\documentclass[final]{IEEEtran}
\pdfoutput=1
\usepackage{amsthm,amssymb,graphicx,multirow,amsmath,color,amsfonts}%,ulem}
\usepackage[update,prepend]{epstopdf}
\usepackage[noadjust]{cite}
\usepackage[latin1]{inputenc}
\usepackage{tikz}
\usepackage{bbm} % for \mathbbm{1}
\usepackage{pdfpages}
\usepackage{tabulary}
\usepackage{multirow}
\usepackage{comment}

\def\nb0{{\mathbf{0}}}
\def\nb1{{\mathbf{1}}}

\newtheorem{lemma}{Lemma}

\newtheorem{definition}{Definition}

\newtheorem{theorem}{Theorem}

\newtheorem{cor}{Corollary}

\newtheorem{remark}{Remark}

\def\argmax{\operatorname{arg~max}}
\allowdisplaybreaks % Allows breaking of eqnarray over multiple pages (avoids unnecessary blanks in the document before eqnarray)

\begin{document}
%\pagenumbering{gobble}
\graphicspath{{./Figures/}}
\title{
%Effective Fading Distribution at the Receiver in a 5G Heterogeneous Cellular Network
%Effective Fading Distribution at the Receiver in a $K$-tier Heterogeneous Cellular Network
Effect of Cell-Selection on the Effective Fading Distribution in a Downlink $K$-tier HetNet
}
\author{
Mustafa A. Kishk and Harpreet S. Dhillon
\thanks{The authors are with Wireless@VT, Department of ECE, Virginia Tech, Blacksburg, VA. Email: \{mkishk, hdhillon\}@vt.edu. The support of the U.S. NSF (Grants CCF-1464293 and CNS-1617896) is gratefully acknowledged.} % remove the date for conference drafts
\vspace{-5ex}}

\maketitle

\begin{abstract}
This paper characterizes the statistics of effective fading gain in multi-tier cellular networks with strongest base station (BS) cell association policy. First, we derive the probability of association with the $n$-th nearest BS in the $k$-th tier. Next, we use this result to derive the probability density function (PDF) of the channel fading gain (effective fading) experienced by the user when associating with the strongest BS. Interestingly, our results show that the effective channel gain distribution solely depends upon the original channel fading and the path-loss exponent. Moreover, we show that in the case of Nakagami-$m$ fading channels (Gamma distribution), the distribution of the effective fading is also Gamma but with a gain of $\frac{\alpha}{2}$ in the shape parameter, where $\alpha$ is the path-loss exponent.%We verify our analysis by showing perfect match with simulation results.
\end{abstract}
\vspace{-3mm}
\section{Introduction} \label{sec:intro}
With the dawn of fifth-generation (5G) cellular systems almost upon us, we are currently witnessing unprecedented changes in the cellular use-cases and consequently the architecture of cellular networks. In the context of this letter, two parallel trends are of particular interest. First, cell sizes are shrinking due to the organic deployment of low-power BSs called small cells~\cite{DhiGanJ2012}. Second, 5G may need to support mobility at speeds upto 500km/hour~\cite{ITUM2015}. These trends together mean that the mobile users may experience orders of magnitude higher handover rates compared to the current networks. This issue has indeed attracted a lot of attention, with cloud radio access networks (C-RANs) and more recently Fog networks presented as possible solutions~\cite{CheChrJ2015}. The occurrence of more frequent handovers necessitates the need to understand the system characteristics {\em during} the handover phase, which we refer to as the {\em cell selection phase}. In this letter, we focus on characterizing the distribution of fading gain experienced by a mobile user during the cell selection phase, which we term as {\em effective fading gain}. Due to the occurence of {\em order statistics} in cell selection, the effective fading gain is shown to deviate significantly from the fading gain observed by the users at other times (when they are not performing cell selection), which is referred to as {\em original fading gain}. It is worth noting that this problem has never been studied before perhaps because it was not quite as meaningful in conventional cellular networks that had far lower handover rates.

Although this exact problem has not been studied before, order statistics have been used to study several problems that have a somewhat similar flavor~\cite{yang2011order}. For instance, in multi-antenna receivers, it is important to exploit the existence of multiple paths by optimally combining them in order to enhance the performance gain. This is usually achieved by giving larger weights to the {\em strongest} paths~\cite{536918,837044,1350921}. Also in multi-user systems, similar procedure appears during scheduling when the transmitter picks the users with {\em strongest} channel state~\cite{1673689,4600211,4012549}. Similar to diversity techniques, selecting stronger users during scheduling necessitates using order statistics in the analysis. Similar decision needs to be made during {\em cell selection} in cellular networks, which is the main focus of this letter. The usual procedure is to measure the received signal strengths from all nearby BSs at the mobile user and then attach it to the BS that provides the strongest signal. While this typically reduces to closest-BS association in conventional single-tier cellular networks, the same is not true in multi-tier HetNets due to different transmit powers across tiers~\cite{DhiGanJ2012}. Assuming same fading gain statistics across all wireless links, in this letter we study the effect of cell selection procedure on the {\em effective} fading gain observed at the mobile user.

\textit{Contributions}. Using tools from stochastic geometry~\cite{AndGupJ2016}, we characterize the PDF of the effective channel fading gain experienced by the typical user in a multi-tier cellular network during cell-selection phase under strongest-BS association policy. As an important intermediate result, we first derive the probability of association of the typical user with the $n$-th nearest BS in the $k$-th tier, which itself is a new and much finer characterization of association probabilities under strongest BS association policy in HetNets. Our results concretely demonstrate that the PDF solely depends upon the distribution of the original fading gain and the path-loss exponent (does not depend upon infrastructure properties, such as the number of BS tiers, their  densities, or their transmission powers). In addition, if we assume independent Nakagami-$m$ fading (Gamma distribution) over all links, the resulting effective fading is also Gamma distributed with the same scale parameter but a gain of $\frac{\alpha}{2}$ in the shape parameter.
%while the shape parameter is higher by a value of $\frac{\alpha}{2}$.

%In this paper we use tools from stochastic geometry to derive the PDF of the channel fading gain experienced by a typical user in a multi-tier cellular network with strongest BS association policy. We first derive the probability of association with the $n$-th nearest BS in the $k$-th tier. Next, we use this result to derive the distribution of the experienced channel fading by the typical user. Our results show that the distribution is indifferent to the number of tiers, their densities, or their transmission power. The results concretely show that the distribution of the channel fading is only function of the path-loss exponent and the distribution of the original fading gain across the communication links. In addition, in the case of Nakagami-$m$ original fading (Gamma distribution) the resulting effective fading is also Gamma distributed with similar scale parameter while the shape parameter is higher by a value of $\frac{\alpha}{2}$.
\vspace{-3mm}
\section{System Model} \label{sec:SysMod}
We consider a $K$-tier cellular network were the locations of BSs in each tier are modeled by an independent Poisson point process (PPP) $\Phi_k=\{x_{k,i}\}\in\mathbb{R}^2$ with density $\lambda_k$. The transmission power of BSs in the $k$-th tier is $P_k$. The locations of all BSs can then be modeled by the PPP $\Psi=\bigcup_{k\in\mathcal{K}}\Phi_k$, where $\mathcal{K}=\{1,2,\dots,K\}$ is the set of indices of the tiers. The locations of the users are modeled by an independent PPP $\Phi_u=\{u_{i}\}\in\mathbb{R}^2$ with density $\lambda_u$. Without loss of generality, the analysis is performed at an arbitrarily chosen user (termed {\em typical} user), whose position can be translated to the origin due to the stationarity of this setup. 
%Without loss of generality (using stationarity of PPP), we focus our analysis on a typical user located at the origin. 
The received signal power at the typical user from a $k$-th tier BS located at $x \in \Phi_k$ is $P_k h_{x}\|x\|^{-\alpha}$, where $h_x$ models fading gain and $\|x\|^{-\alpha}$ represents standard power-law pathloss with exponent $\alpha>2$. % and $h_{x}$ models the cumulative effect of both slow (shadowing) and fast fading. 
To differentiate it from the {\em effective} fading gain studied in the next Section, we refer to $h_{x}$ as the {\em original} fading gain. For notational ease, the set of all original fading gains $\{h_x\}$, $x\in \Phi_k$, $\forall k \in \mathcal{K}$ over all links between the typical user and the BSs in the network is denoted by $\mathcal{H}$. 
%The set of original fading gains across all links between the typical user and the BSs is $\mathcal{H}$. 
It is assumed that $h_x$ is independent and identically distributed across all links with PDF $f_h(h)$ and cumulative distribution function (CDF) $F_h(h)$. While all the results will be derived for this general distribution, we will also specialize them to Nakagami-$m$ fading case for providing insights and numerical comparisons. In this case, we have ${f}_{h}(y)= \left(\frac{m}{\Omega}\right)^m\frac{y^{m-1}}{\Gamma(m)} \exp(-\frac{m}{\Omega}y), y\geq 0$, which is Gamma distribution with scale parameter $\frac{\Omega}{m}$ and shape parameter $m$ (i.e. $h\sim \textnormal{Gamma}\left(\frac{\Omega}{m},m\right)$), where $\Omega$ is the mean of $h$ (usually assumed to be $1$ in Nakagami-$m$ fading channels). Also $F_h(y)= 1- \frac{\Gamma(m,\frac{m}{\Omega}y)}{\Gamma(m)}$. For cell selection, we assume that the typical user connects to the BS that provides maximum received power. For this setup, we define effective fading gain experienced at the typical user as follows. 
%Each user is assumed to associate with the BS that provides maximum instantaneous received power ({\em strongest} BS). The received power by the typical user if it connects to a BS in the $k$-th tier located at $x$ will be $P_k h_{x}\|x\|^{-\alpha}$, where $h_x$ is the channel fading gain, and $\|x\|^{-\alpha}$ represents the path-loss with exponent $\alpha>2$. Note that $h_x$ is the resultant gain of both slow and fast fading. A common assumption in literature is to neglect the small fading (e.g. shadowing) and model the fast fading using the Nakagami-$m$ model. 
%We assume independent fading gains among all links with general PDF $f_h(h)$ and CDF $F_h(h)$. We will refer to the distribution of $h$ in the rest of the paper as the {\em original} fading distribution. In the following section, we derive the distribution of the effective fading gain, which is defined next.
\begin{definition}[Effective Fading Gain] \label{def:1}
The effective fading gain $h^*$ represents the fading gain experienced by the typical user during cell-selection phase when associating with the BS located at $x^*$, where 
%. The location $x^*$ can be mathematically represented as follows:
\begin{align}
x^*=\underset{x \in \Phi_k, k \in \mathcal{K}}{\argmax}\ P_k h_x\|x\|^{-\alpha}.
%\arg\underset{j\in\mathcal{K}}{\max}\ \underset{x\in\Phi_{\rm j}}{\max}\ P_j h_x\|x\|^{-\alpha}.
\label{eq:cell_selection}
\end{align}
\end{definition}
Note that, in principle, $h_x$ can be interpreted as the cumulative fading gain incorporating both slow and fast fading effects. This generality is mathematically provided by the general distribution of $h$. Also note that the cell-selection policy can be specialized to both instantaneous and average power-based cell selection by choosing appropriate distribution $f_h(\cdot)$. 
\section{Distribution of Effective Fading Gain}
Since fading gain appears in the cell selection criterion given by Eq. \ref{eq:cell_selection}, the distribution of the effective fading gain $h^*$ experienced by the typical user will, in general, be different from the distribution of the original fading gain $h$ (due to the effect of order statistics). Before starting the derivation of the PDF of $h^*$, we introduce an indicator function $\delta_{k,n}$ for which $\delta_{k,n}=1$ if the $n$-th nearest BS in the $k$-th tier is the serving BS and $\delta_{k,n}=0$ otherwise. This function can be expressed as
\begin{align}
\label{Eq:2}
\delta_{k,n}&=\prod_{x_k\in\Phi_k\backslash x_{k,n}}\mathbbm{1}(\|x_{k,n}\|^{-\alpha}h_{k,n}\geq\|x_k\|^{-\alpha}h_{x_k})\\ &\times\prod_{j\in\mathcal{K}\backslash k}\prod_{x_j\in\Phi_j}\mathbbm{1}\left(P_k\|x_{k,n}\|^{-\alpha}h_{k,n}\geq P_j\|x_j\|^{-\alpha}h_{x_j}\right),\nonumber
\end{align}
where $\mathbbm{1}(\Xi)$ is an indicator function that equals to 1 when the condition $\Xi$ is satisfied and equals to zero otherwise, and $h_{k,n}$ is the original fading gain for the channel between the typical user and the $n$-th nearest BS in the $k$-th tier. Using the above function, the CDF of $h^*$ is defined as follows:
%To provide the CDF of the effective fading gain $h^*$ we use the conditional probability derived in Eq.~\ref{Eq_lemma1_final2} in Appendix~\ref{app:assoc} as follows:
\begin{align}
\label{Eq_pre_lemma_1}
\mathbb{P}(h^*\leq y)=&\sum_{k\in\mathcal{K}}\sum_{n=1}^{\infty}\mathbb{P}(h_{k,n}\leq y,\delta_{k,n}=1)\nonumber\\
=&\sum_{k\in\mathcal{K}}\sum_{n=1}^{\infty}\mathbb{E}_{\Psi,\mathcal{H}}\left[\mathbbm{1}(h_{k,n}\leq y)\delta_{k,n}\right]\nonumber\\
\stackrel{(a)}{=}&\sum_{k\in\mathcal{K}}\sum_{n=1}^{\infty}\mathbb{E}_{h_{k,n}}\left[\mathbbm{1}(h_{k,n}\leq y)\mathbb{P}(\delta_{k,n}=1|h_{k,n})\right]\nonumber\\
\stackrel{(b)}{=}&\sum_{k\in\mathcal{K}}\sum_{n=1}^{\infty}\int_0^{y}f_h(h_{k,n})P_{(k,n)|h_{k,n}}{\rm d}h_{k,n},
\end{align}
where $(a)$ follows from the fact that $\mathbbm{1}(h_{k,n}\leq y)$ and $\delta_{k,n}$ are conditionally independent, conditioned on $h_{k,n}$, and $P_{(k,n)|h_{k,n}}$ in $(b)$ is the probability of association with the $n$-th nearest BS in the $k$-th tier conditioned on $h_{k,n}$. Hence, we first derive an expression for $P_{(k,n)|h_{k,n}}$ in Lemma~\ref{lem:assoc}, using which we will derive the PDF of $h^*$ in Theorem~\ref{thm:effective_general}.
\subsection{Conditional Association Probability}
We first derive an expression for $P_{(k,n)|h_{k,n}}$ in Lemma~\ref{lem:assoc}. 
%We now derive the probability of association of the typical user with the $n$-th nearest BS in the $k$-th tier conditioned on $h_{k,n}$.
\begin{lemma}[Association Probability]\label{lem:assoc}
The conditional probability of the typical user associating with the $n$-th nearest BS in the $k$-th tier is:
%\begin{align}
%\label{Eq_lemma1_1}
%P_{(k,n)\big|h_{k,n}}&= \left(\frac{1}{g_2(h_{k,n})+1}\right)^ng_1(h_{k,n})^{n-1},
%\end{align}
%While the unconditional probability is:
\begin{align}
\label{Eq_lemma1_2}
P_{(k,n)|h_{k,n}}=\left(\frac{1}{g_2(h_{k,n})+1}\right)^ng_1(h_{k,n})^{n-1},
\end{align}
where $g_1(h_{k,n})=\frac{2}{\alpha}\int_{0}^{1}F_h\left(yh_{k,n}\right)y^{\frac{2}{\alpha}-1}{\rm d}y$, $g_2(h_{k,n})=\frac{2}{\alpha}\int_{1}^{\infty}\bar{F}_h\left(yh_{k,n}\right)y^{\frac{2}{\alpha}-1}{\rm d}y+\Big(\frac{2}{\mathcal{B}_k\alpha}\int_{0}^{\infty}\bar{F}_h\left(yh_{k,n}\right)y^{\frac{2}{\alpha}-1}{\rm d}y\Big)$, ${\mathcal{B}_k}=\frac{{\lambda_k}{P_k}^{\frac{2}{\alpha}}}{\sum_{j\in\mathcal{K}\backslash k}{\lambda_j}\left({P_j}\right)^{\frac{2}{\alpha}}}$, and $\bar{F}_h(h)=1-F_h(h)$.   
\end{lemma}
\begin{IEEEproof}
See Appendix~\ref{app:assoc}.
\end{IEEEproof} 
\begin{remark} \label{rem:Bias_term}
Note that the effect of BS density and their transmit powers appears in the term ${\mathcal{B}_k}$ in $g_2(h_{k,n})$ in Eq.~\ref{Eq_lemma1_2}. Consistent with the intuition, the conditional association probability is an increasing function of both $\lambda_k$ and $P_k$. %This means that increasing the transmit power or the density of a given tier will lead to higher probability that the serving BS belongs to that tier. 
%Note that the association to the $n$-th nearest BS in the $k$-th tier is biased through the term ${\mathcal{B}_k}$ which appears in the expression of $g_2(h_{k,n})$ in Eq.~\ref{Eq_lemma1_2}. We can note that either the density or the transmission power of the $k$-th tier are directly proportional to the bias term $\mathcal{B}_k$ and consequently, directly proportional to $P_{(k,n)}$. This note agrees with intuition, as increasing the transmission power or the density of the $k$-th tier leads to higher probability that the serving BS belongs to this tier. 
\end{remark}
For the case of Nakagami-$m$ original fading, the integrals in $g_1(h_{k,n})$ and $g_2(h_{k,n})$ can be reduced to closed forms.
\begin{cor}%[Association Probability with Nakagami-$m$ Fading]\label{cor:assoc_nakagami}
In the case of Nakagami-$m$ original fading, $g_1(h_{k,n})$ and $g_2(h_{k,n})$ can be simplified as follows:
\begin{align}
g_1(h_{k,n})&=1-\frac{\left(\frac{\Omega}{mh_{k,n}}\right)^{\frac{2}{\alpha}}\gamma\left(m+\frac{2}{\alpha},\frac{mh_{k,n}}{\Omega}\right)+\Gamma\left(m,\frac{mh_{k,n}}{\Omega}\right)}{\Gamma(m)}\nonumber\\
g_2(h_{k,n})&=\frac{\left(\frac{\Omega}{mh_{k,n}}\right)^{\frac{2}{\alpha}}\left(\Gamma\left(m+\frac{2}{\alpha},\frac{mh_{k,n}}{\Omega}\right)+\frac{\Gamma(m+\frac{2}{\alpha})}{\mathcal{B}_k}\right)}{\Gamma(m)}\nonumber\\
&-\frac{\Gamma\left(m,\frac{mh_{k,n}}{\Omega}\right)}{\Gamma(m)},
\end{align}
where $\Gamma(a,b)$, and $\gamma(a,b)$ are the upper and lower incomplete Gamma functions, respectively.
\end{cor}
\subsection{Effective Fading Distribution Analysis}
In the following Theorem we provide the PDF of the effective fading gain $h^*$ in terms of general original fading.
\begin{theorem}[Effective Fading Distribution with General Fading] \label{thm:effective_general}
The PDF of the effective fading gain in a $K$-tier network under general cell-selection policy given by Definition~\ref{def:1} is: 
\begin{align}
\label{Eq_theorem1_1}
f_{h^*}(y)=\frac{\frac{\alpha}{2} {y}^{\frac{2}{\alpha}} f_{h}(y)}{\int_{0}^{\infty}\bar{F}_h(z)z^{\frac{2}{\alpha}-1}{\rm d}z}.
\end{align}
\end{theorem}
\begin{IEEEproof}
See Appendix~\ref{app:effective}.
\end{IEEEproof}
\begin{remark} \label{rem:alpha_fading_dist}
Note that in Eq.~\ref{Eq_theorem1_1} the distribution of the effective channel fading gain is agnostic of the number of tiers, the transmission powers, and the densities of the tiers. Interestingly, the distribution only depends on the original fading distribution and the path-loss exponent $\alpha$. %This result can be very useful for cellular network system designers, specifically in the design of transceivers where channel statistics play an important role.
\end{remark}
In the case of Nakagami-$m$ fading, the distribution of the effective fading gain is presented in the following corollary.
\begin{cor}[Effective Fading Distribution with Nakagami-$m$ Original Fading]\label{cor:effective_nakagami}
In the case of Nakagami-$m$ original fading with $h\sim \textnormal{Gamma}\left(\frac{\Omega}{m},m\right)$, we have
\begin{align}
\label{Eq_cor2_1}
f_{h^*}(y)=\left(\frac{m}{\Omega}\right)^{m+\frac{2}{\alpha}}y^{\frac{2}{\alpha}+m-1}\frac{\exp(-\frac{m}{\Omega}y)}{{\Gamma(m+\frac{2}{\alpha})}}.
\end{align}
\end{cor}
\begin{remark} \label{rem:m_effect}
From Corollary \ref{cor:effective_nakagami}, we note that in the case of Nakagami-$m$ original fading $h\sim \textnormal{Gamma}\left(\frac{\Omega}{m},m\right)$, the distribution of the effective fading gain is also Nakagami-$m$ but with a different shape parameter: $h^*\sim \textnormal{Gamma}(\frac{\Omega}{m},m+\frac{2}{\alpha})$. The $\frac{2}{\alpha}$ gain in the shape parameter represents the decrease in the severity of the effective fading gain distribution as a result of base station selection. It is also noteworthy that this gain gets negligible as the value of $m$ gets larger.
\end{remark}
\begin{remark}
\label{rem:SysDesign}
The reduction in the perceived severity of the fading at the receiver (discussed in Remark~\ref{rem:m_effect}) can be used to lower the transmission power of the BS for about one channel coherence time immediately after the cell selection phase. This can be argued using multiple performance metrics. For instance, due to reduction in the channel fluctuations, one needs a relatively lower {\em fade margin} during this time, which directly translates to lower transmit power at the BS. Similarly, due to less severe fading, the {\em bit error rate} will decrease, which means the BS can transmit at a lower power to achieve the same target bit error rate as before.
\end{remark}
\vspace{-3ex}
\subsection{Simulation Results} 
\begin{figure}
\centering
\includegraphics[width=0.7\columnwidth]{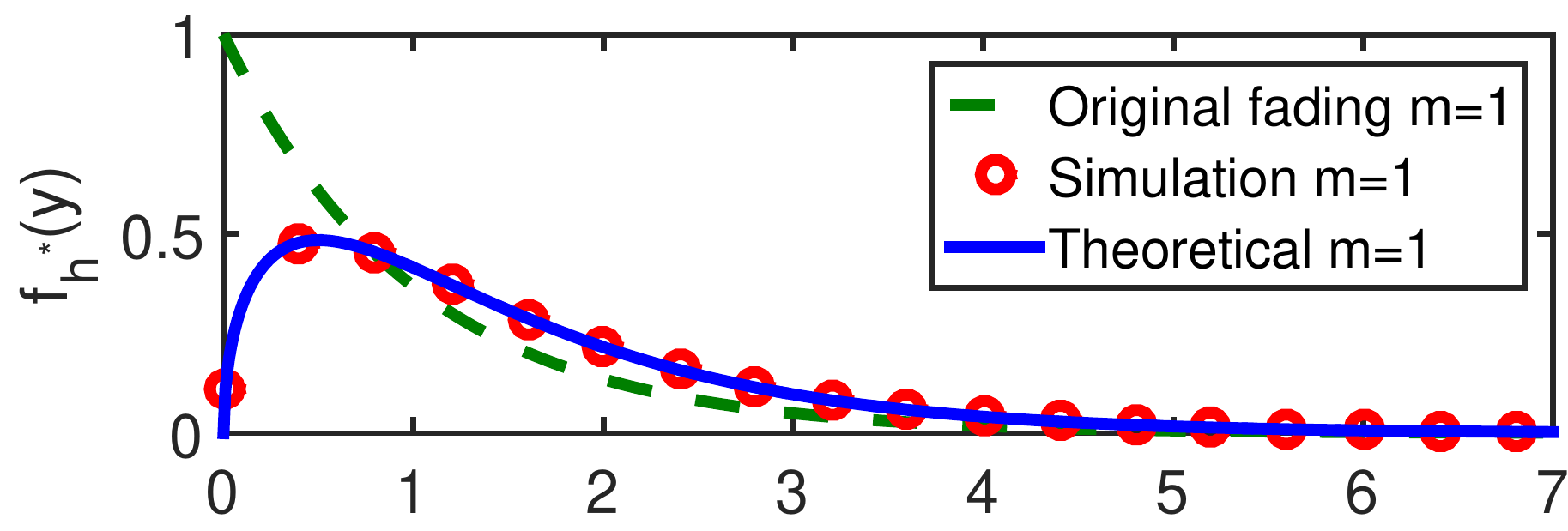}
\includegraphics[width=0.7\columnwidth]{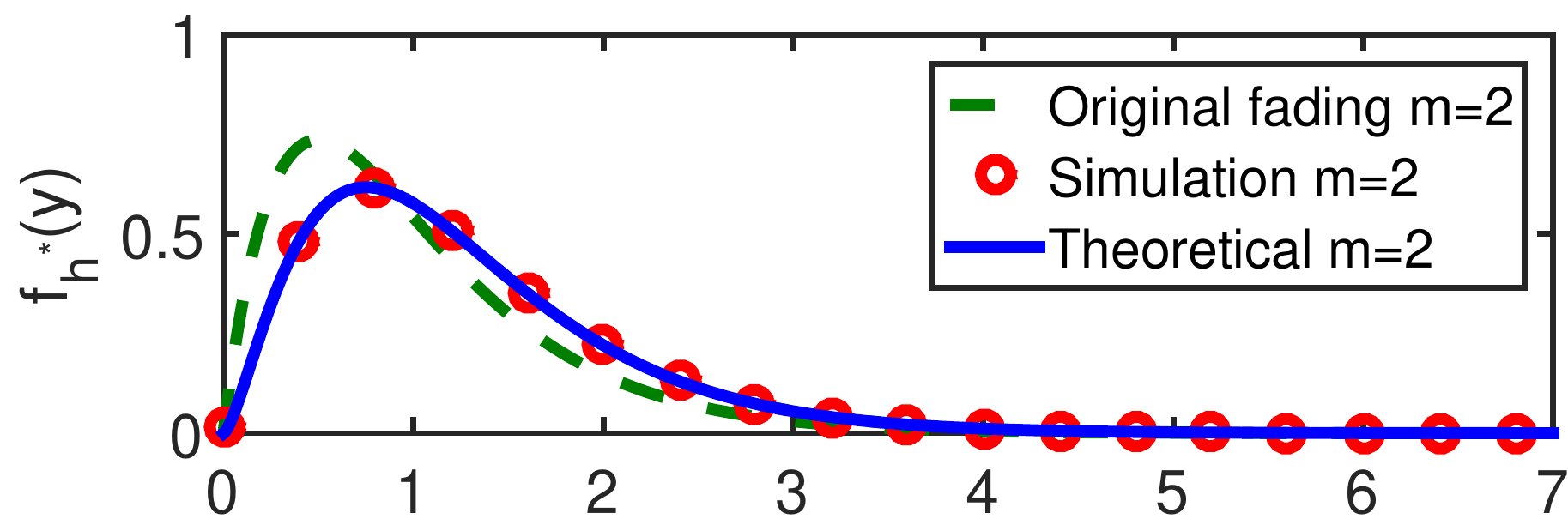}
\includegraphics[width=0.7\columnwidth]{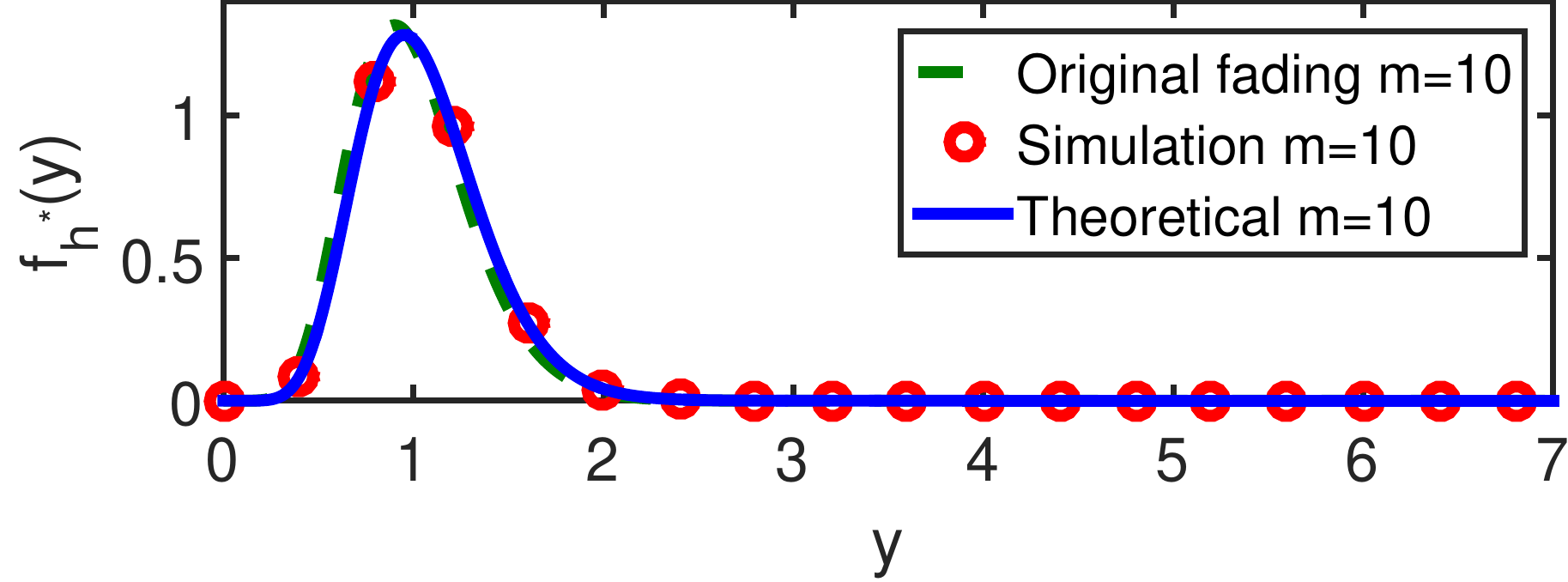}
\caption{PDF of both original channel fading gain and effective channel fading gain for Nakagami-$m$ original fading case with different values of $m$ ($P_1=1$, $P_2=2$, $\lambda_1=1$, and $\lambda_2=2$). Results are independent of $K$.
}
\label{fig}
\vspace{-3mm}
\end{figure}
%Here we evaluate our analytical results by comparing them with simulation. 
In Fig.~\ref{fig}, we present the PDF of the effective fading gain $h^*$ for the case of Nakagami-$m$ original fading with $\Omega=1$ for several values of $m$. Along with the analytical results, the simulation results are also provided for $K=2$ tiers. As discussed already in Remark~\ref{rem:alpha_fading_dist}, these results do not depend upon the number of tiers (value of $K$), which was also verified separately with simulations.  
%We also note that our analytical results perfectly match with the simulations, which validates our analysis. 
By comparing these distributions with the corresponding original fading distributions, we deduce that the PDF of $h^*$ almost coincides with the PDF of the original fading $h$ as the value of $m$ is increased, whereas they are significantly different for smaller values of $m$.  This is consistent with insights gained in Remark~\ref{rem:m_effect}.
%We note that our analytical results are perfectly matching with simulations. In addition, as noted in Remark~\ref{rem:m_effect}, as the value of $m$ increases the PDF of $h^*$ almost coincides with the PDF of the original fading $h$.
\vspace{-3mm}
\section{Conclusion}
%Owing to network densification, 5G networks may experience orders of magnitude higher handover rates. It is therefore important to understand network properties during the handover phases. 
In this letter, we derived the PDF of the effective fading gain observed by the typical user during the handover phase in a $K$-tier HetNet under strongest-BS association policy. As an intermediate result  (which is in fact important in its own right), we provided a much finer characterization of association probabilities by deriving the probability that the typical user is served by the $n$-th nearest BS in the $k$-th tier. Our results concretely demonstrate that the effective channel fading distribution during cell selection phase is significantly different from the original fading distribution. In fact, association with the strongest BS reduces perceived {\em severity} of fading in general due to order statistics. 
One useful consequence of this result is the possibility of lowering the transmit power of the serving BS to the typical user for about one channel coherence time immediately after the cell selection phase.%
\vspace{-5mm}
%In this letter, we derived the PDF of the effective fading gain for a $K$-tier network with strongest BS association policy. Different from the location based association policies (e.g nearest BS association) where the effective fading distribution is similar to the original fading distribution, our results show a decrease in the severity of the channel fading due to associating with the strongest BS. While we assumed PPP for the locations of BSs, a good extension to this work is to study the PDF of effective fading for the case of different point process.
\appendix
\subsection{Proof of Lemma~\ref{lem:assoc}}\label{app:assoc}
The association probability conditioned on $h_{k,n}$ can be derived as follows:
\begin{align}
\label{Eq_lemma_1_proof}
&P_{(k,n)|h_{k,n}}=\mathbb{E}_{x_{k,n}}\mathbb{E}_{\Phi_k}^{{!x_{k,n}}}\mathbb{E}_{\Psi\backslash\Phi_k}\mathbb{E}_{\mathcal{H}}\Bigg[\prod_{x_k\in\Phi_k\backslash x_{k,n}}\mathbbm{1}(\|x_{k,n}\|^{-\alpha}h_{k,n}\nonumber\\ &\geq\|x_k\|^{-\alpha}h_{x_k})\times\prod_{j\in\mathcal{K}\backslash k}\prod_{x_j\in\Phi_j}\mathbbm{1}\left(P_k\|x_{k,n}\|^{-\alpha}h_{k,n}\geq P_j\|x_j\|^{-\alpha}h_{x_j}\right)\Bigg]\nonumber\\
&\stackrel{(a)}{=}\mathbb{E}_{x_{k,n}}\mathbb{E}_{\Phi_k}^{{!x_{k,n}}}\mathbb{E}_{\Psi\backslash\Phi_k}\Bigg[\prod_{x_k\in\Phi_k\backslash x_{k,n}}\mathbb{P}(\|x_{k,n}\|^{-\alpha}h_{k,n}\geq\|x_k\|^{-\alpha}h_{x_k})\nonumber\\ &\times\prod_{j\in\mathcal{K}\backslash k}\prod_{x_j\in\Phi_j}\mathbb{P}\left(P_k\|x_{k,n}\|^{-\alpha}h_{k,n}\geq P_j\|x_j\|^{-\alpha}h_{x_j}\right)\Bigg]\nonumber\\
&\stackrel{(b)}{=}\mathbb{E}_{x_{k,n}}\mathbb{E}_{\Phi_k}^{{!x_{k,n}}}\mathbb{E}_{\Psi\backslash\Phi_k}\Bigg[\prod_{x_k\in\Phi_k\backslash x_{k,n}}F_h\left(\frac{\|x_k\|^{\alpha}}{\|x_{k,n}\|^{\alpha}}h_{k,n}\right)\nonumber\\ &\times\prod_{j\in\mathcal{K}\backslash k}\prod_{x_j\in\Phi_j}F_h\left(\frac{P_k\|x_j\|^{\alpha}}{P_j\|x_{k,n}\|^{\alpha}}h_{k,n}\right)\Bigg]\nonumber\\
&\stackrel{(c)}{=}\mathbb{E}_{x_{k,n}}\Bigg[
\underbrace{\mathbb{E}_{\Phi_k}^{{!x_{k,n}}}\Bigg[\prod_{x_k\in\Phi_k\backslash x_{k,n}}F_h\left(\frac{\|x_k\|^{\alpha}}{\|x_{k,n}\|^{\alpha}}h_{k,n}\right)\Bigg]}_{I_1(x_{k,n})}\nonumber\\ 
&\times \underbrace{\prod_{j\in\mathcal{K}\backslash k}\mathbb{E}_{\Phi_j}\Bigg[\prod_{x_j\in\Phi_j}F_h\left(\frac{P_k\|x_j\|^{\alpha}}{P_j\|x_{k,n}\|^{\alpha}}h_{k,n}\right)\Bigg]}_{I_2(x_{k,n})}
\Bigg] \nonumber \\
&=\mathbb{E}_{x_{k,n}}\left[I_1(x_{k,n})I_2(x_{k,n})\right],
\end{align}
where ($a$) is due to the independence of fading gain across all links, ($b$) follows by substituting for the CDF of the original fading gain $F_h(h)$, and step ($c$) follows from the assumption that the $K$-tiers are modeled by {\em independent} PPPs. %Now let $P_{(k,n)|h_{k,n}}=\mathbb{E}_{x_{k,n}}\left[I_1(x_{k,n})I_2(x_{k,n})\right]$ where
%\begin{align}
%I_1(x_{k,n})&=\mathbb{E}_{\Phi_k}^{{!x_{k,n}}}\Bigg[\prod_{x\in\Phi_k\backslash x_{k,n}}F_h\left(\frac{|x\|^{\alpha}}{\|x_{k,n}\|^{\alpha}}h_{k,n}\right)\Bigg]\nonumber\\
%I_2(x_{k,n})&=\prod_{j\in\mathcal{K}\backslash k}\mathbb{E}_{\Phi_j}\Bigg[\prod_{x\in\Phi_j}F_h\left(\frac{P_k\|x\|^{\alpha}}{P_j\|x_{k,n}\|^{\alpha}}h_{k,n}\right)\Bigg]
%\end{align}

The term $I_1(x_{k,n})$ can be derived by splitting $\Phi_k$ into $\Phi_k\cap B(0,\|x_{k,n}\|)$ and $\Phi_k\cap B(0,\|x_{k,n}\|)^c$, where $B(0,\|x_{k,n}\|)$ is the ball of radius $\|x_{k,n}\|$ centered at the origin, and $B(0,\|x_{k,n}\|)^c$ is its compliment. Note that conditioned on the location of the $n$-th nearest BS $x_{k,n}$, the number of points inside the ball $\Phi_k\cap B(0,\|x_{k,n}\|)$ is $n-1$, and their locations are uniformly distributed (follows from the definition of PPP). Hence, $I_1(x_{k,n})$ can be derived as follows:
\begin{align}
\label{Eq_I_1}
&I_1(x_{k,n})=\mathbb{E}_{\Phi_k}^{{!x_{k,n}}}\Bigg[\prod_{x\in\Phi_k\cap B(0,\|x_{k,n}\|)}F_h\left(\frac{\|x\|^{\alpha}}{\|x_{k,n}\|^{\alpha}}h_{k,n}\right)\Bigg]\nonumber\\&\times\mathbb{E}_{\Phi_k}^{{!x_{k,n}}}\Bigg[\prod_{x\in\Phi_k\cap B(0,\|x_{k,n}\|)^c}F_h\left(\frac{\|x\|^{\alpha}}{\|x_{k,n}\|^{\alpha}}h_{k,n}\right)\Bigg]\nonumber\\
&\stackrel{(d)}{=}\left(\frac{1}{\pi\|x_{k,n}\|^2}\int_{x\in\mathbb{R}^2\cap B(0,\|x_{k,n}\|)}F_h\left(\frac{\|x\|^{\alpha}}{\|x_{k,n}\|^{\alpha}}h_{k,n}\right){\rm d}x\right)^{n-1}\nonumber\\&\times\exp\left(-\lambda_k\int_{x\in\mathbb{R}^2\cap B(0,\|x_{k,n}\|)^c}1-F_h\left(\frac{\|x\|^{\alpha}}{\|x_{k,n}\|^{\alpha}}h_{k,n}\right){\rm d}x\right)\nonumber\\
&\stackrel{(e)}{=}\left(\frac{2}{\alpha}\int_{0}^{1}F_h\left(yh_{k,n}\right)y^{\frac{2}{\alpha}-1}{\rm d}y\right)^{n-1}\nonumber\\&
\times\exp\left(-\frac{2\pi\lambda_kr_{k,n}^2}{\alpha}\int_{1}^{\infty}\bar{F}_h\left(yh_{k,n}\right)y^{\frac{2}{\alpha}-1}{\rm d}y\right),
\end{align}
where $\bar{F}_h(h)=1-F_h(h)$. The first term in ($d$) is due to the uniform distribution of the points inside $\Phi_k\cap B(0,\|x_{k,n}\|)$, the second term in ($d$) follows by applying PGFL of PPP~\cite{haenggi2012stochastic}. Note that the point process $\Phi_k\cap B(0,\|x_{k,n}\|)^c$ remains a PPP due to Slyvniak's theorem. Step ($e$) follows from converting to polar coordinates where $r_{k,n}=\|x_{k,n}\|$ and simple manipulations to the integrals. The term $I_2(x_{k,n})$ can also be derived using PGFL of PPP and similar procedure to steps (d) and (e) which results in:
\begin{align}
\label{Eq_I_2}
&I_2(x_{k,n})=\prod_{j\in\mathcal{K}\backslash k}\exp\left(-2\pi\lambda_j\int_0^{\infty}\bar{F}_h\left(\frac{P_kr^{\alpha}}{P_jr_{k,n}^{\alpha}}h_{k,n}\right)r{\rm d}r\right)\nonumber\\
&=\prod_{j\in\mathcal{K}\backslash k}\exp\left(-\frac{2\pi\lambda_jr_{k,n}^2}{\alpha}\left(\frac{P_j}{P_k}\right)^{\frac{2}{\alpha}}\int_0^{\infty}\bar{F}_h\left(yh_{k,n}\right)y^{\frac{2}{\alpha}-1}{\rm d}y\right)\nonumber\\
&=\exp\left(-\frac{2\pi\lambda_kr_{k,n}^2}{\alpha\mathcal{B}_k}\int_0^{\infty}\bar{F}_h\left(yh_{k,n}\right)y^{\frac{2}{\alpha}-1}{\rm d}y\right),
\end{align}
where ${\mathcal{B}_k}=\frac{{\lambda_k}{P_k}^{\frac{2}{\alpha}}}{\sum_{j\in\mathcal{K}\backslash k}{\lambda_j}\left({P_j}\right)^{\frac{2}{\alpha}}}$. Now substitute Eq.~\ref{Eq_I_1},~\ref{Eq_I_2} in Eq.~\ref{Eq_lemma_1_proof} , which gives
\begin{align}
\label{Eq_lemma1_final}
&P_{(k,n)|h_{k,n}}=\mathbb{E}_{r_{k,n}}\left[g_1(h_{k,n})^{n-1}\exp\left(-\lambda_k\pi r_{k,n}^2g_2(h_{k,n})\right)\right]\nonumber\\
&=\int_{0}^{\infty}f_{r_{k,n}}(r)g_1(h_{k,n})^{n-1}\exp\left(-\lambda_k\pi r^2g_2(h_{k,n})\right){\rm d}r
\end{align}
where $g_1(h_{k,n})=\frac{2}{\alpha}\int_{0}^{1}F_h\left(yh_{k,n}\right)y^{\frac{2}{\alpha}-1}{\rm d}y,\ g_2(h_{k,n})=\frac{2}{\alpha}\int_{1}^{\infty}\bar{F}_h\left(yh_{k,n}\right)y^{\frac{2}{\alpha}-1}{\rm d}y+\frac{2}{\alpha\mathcal{B}_k}\int_{0}^{\infty}\bar{F}_h\left(yh_{k,n}\right)y^{\frac{2}{\alpha}-1}{\rm d}y$. Substituting $f_{r_{k,n}}(r)=\frac{2}{\Gamma(k)}(\lambda_k\pi)^k r^{(2k-1)}\exp(-\lambda_k\pi r^2)$~\cite{haenggi2012stochastic} in Eq.~\ref{Eq_lemma1_final} with some algebraic manipulations leads to the final result in Lemma~\ref{lem:assoc}.
%\begin{align}
%\label{Eq_lemma1_final2}
%&P_{(k,n)|h_{k,n}}=\left(\frac{1}{g_2(h_{k,n})+1}\right)^ng_1(h_{k,n})^{n-1}.
%\end{align}
%The result in Lemma~\ref{lem:assoc} follows by %deconditioning Eq.~\ref{Eq_lemma1_final2} on $h_{k,n}$ by 
%integrating over the PDF of the original fading $f_h(h_{k,n})$.
%\subsection{Proof of Corollary~\ref{cor:assoc_nakagami}}\label{app:assoc_nakagami}
\vspace{-3mm}
\subsection{Proof of Theorem~\ref{thm:effective_general}}\label{app:effective}
Using Eq.~\ref{Eq_pre_lemma_1} and the result in Lemma~\ref{lem:assoc}, the CDF of the effective fading gain $h^*$ can be derived as follows:
\begin{align}
\label{Eq_thm_app_1}
\mathbb{P}(h^*\leq y) =&\sum_{k\in\mathcal{K}}\int_0^{y}f_h(h_{k,n})\sum_{n=1}^{\infty}\left(P_{(k,n)|h_{k,n}}\right){\rm d}h_{k,n}\\
\stackrel{(f)}{=}&\sum_{k\in\mathcal{K}}\int_0^{y}f_h(h_{k,n})\frac{1}{1+g_2(h_{k,n})-g_1(h_{k,n})}{\rm d}h_{k,n},\nonumber
\end{align}
where step ($f$) comes from using the summation of a geometric series. It can easily be shown with simple algebraic manipulations that ${1+g_2(h_{k,n})-g_1(h_{k,n})}=\left(\frac{1}{\tilde{\mathcal{B}}_k}\right)\frac{2}{\alpha}h_{k,n}^{-\frac{2}{\alpha}}\int_0^{\infty}\bar{F}_h(z)z^{\frac{2}{\alpha}-1}{\rm d}z$, where $\tilde{\mathcal{B}}_k=\frac{\mathcal{B}_k}{\mathcal{B}_k+1}$. In this expression, we note that $\tilde{\mathcal{B}}_k$ is the only term that is function of $k$. Moreover, note that $\tilde{\mathcal{B}}_k$ is not function of $h_{k,n}$. Hence, the summation in Eq.~\ref{Eq_thm_app_1} can be handled as follows:
\begin{align}
\label{Eq_thm_app_2}
\mathbb{P}(&h^*\leq y)=\left(\int_0^{y}f_h(h_{k,n})\frac{\alpha h_{k,n}^{\frac{2}{\alpha}}}{2\int_0^\infty\bar{F}(z)z^{\frac{2}{\alpha}-1}{\rm d}z}{\rm d}h_{k,n}\right)\sum_{k\in\mathcal{K}}\tilde{\mathcal{B}}_k\nonumber\\
&\stackrel{(g)}{=}\left(\int_0^{y}f_h(h_{k,n})\frac{\alpha h_{k,n}^{\frac{2}{\alpha}}}{2\int_0^\infty\bar{F}(z)z^{\frac{2}{\alpha}-1}{\rm d}z}{\rm d}h_{k,n}\right),
\end{align}   
where step ($g$) is due to $\sum_{k\in\mathcal{K}}\tilde{\mathcal{B}}_k=1$. Taking the derivative of this CDF w.r.t $y$, the final result in Theorem~\ref{thm:effective_general} follows. 
\vspace{-2mm}
%\subsection{Proof of Corollary~\ref{cor:effective_nakagami}}\label{app:effective_nakagami}
\bibliographystyle{IEEEtran}
\bibliography{Dhillon_WCL2017-0261.bbl}

\end{document}